\documentclass[runningheads]{llncs}

\usepackage{eccv}

\usepackage{eccvabbrv}

\usepackage{graphicx}
\usepackage{booktabs}

\usepackage[accsupp]{axessibility}  

\usepackage{multirow}
\usepackage{booktabs}
\usepackage[table]{xcolor} 
\usepackage{adjustbox}

\usepackage[pagebackref,breaklinks,colorlinks,citecolor=eccvblue]{hyperref}

\usepackage{orcidlink}

\begin{document}

\title{RT-GS: Gaussian Splatting with Reflection and Transmittance Primitives}


\author{Kunnong Zeng\inst{1}\orcidlink{0009-0002-1976-6340} \and
Chensheng Peng\inst{2}\orcidlink{0000-0001-9213-5970} \and
Yichen Xie\inst{2}\orcidlink{0000-0003-4443-8795} \and
Masayoshi Tomizuka\inst{2}\orcidlink{0000-0003-0206-6639} \and
Cem Yuksel\inst{1}\orcidlink{0000-0002-0122-4159}}

\authorrunning{K. Zeng et al.}

\institute{University of Utah \and
University of California, Berkeley}

\maketitle

\begin{abstract}
Gaussian Splatting is a powerful tool for reconstructing diffuse scenes, but it struggles to simultaneously model specular reflections and the appearance of objects behind semi-transparent surfaces. These specular reflections and transmittance are essential for realistic novel view synthesis, and existing methods do not properly incorporate the underlying physical processes to simulate them. To address this issue, we propose RT-GS, a unified framework that integrates a microfacet material model and ray tracing to jointly model specular reflection and transmittance in Gaussian Splatting. We accomplish this by using separate Gaussian primitives for reflections and transmittance, which allow modeling distant reflections and reconstructing objects behind transparent surfaces concurrently. We utilize a differentiable ray tracing framework to obtain the specular reflection and transmittance appearance. Our experiments demonstrate that our method successfully produces reflections and recovers objects behind transparent surfaces in complex environments, achieving significant qualitative improvements over prior methods where these specular light interactions are prominent.


  \keywords{Novel view synthesis \and Gaussian splatting \and Ray tracing}
\end{abstract}

\section{Introduction}
\label{sec:intro}
Reconstructing 3D scenes from multi-view input 2D images and rendering images for novel viewpoints is a fundamental problem in novel view synthesis. Neural Radiance Fields (NeRF) \cite{NeRF} achieve good reconstruction quality by leveraging the expressive power of neural networks. However, the high computational cost of NeRFs makes it difficult to achieve real-time performance. Recently, 3D Gaussian Splatting (3D-GS) \cite{3dgs} has demonstrated fast reconstruction and rendering. However, 3D-GS primarily relies on rasterization, which is well-suited for diffuse appearance but fundamentally limited in modeling multi-bounce light transport and view-dependent specular effects such as reflections. 

Several recent works extend GS with additional mechanisms for non-diffuse phenomena. EnvGS \cite{envgs} utilizes ray tracing on separate environment Gaussian primitives to model specular reflection. While it can synthesize near-field reflections, it still struggles to reconstruct complex reflections and dielectric surfaces due to its simplified rendering functions. Moreover, it does not consider transmittance of semi-transparent surfaces. TransparentGS \cite{transparentgs} integrates transparent Gaussian primitives to model transparent objects. However, it bakes the environment around the transparent objects to generate refraction effects and fails to reconstruct objects 
behind semi-transparent surfaces.

\begin{figure}[tb]
    \centering
    \includegraphics[width=1.0\linewidth]{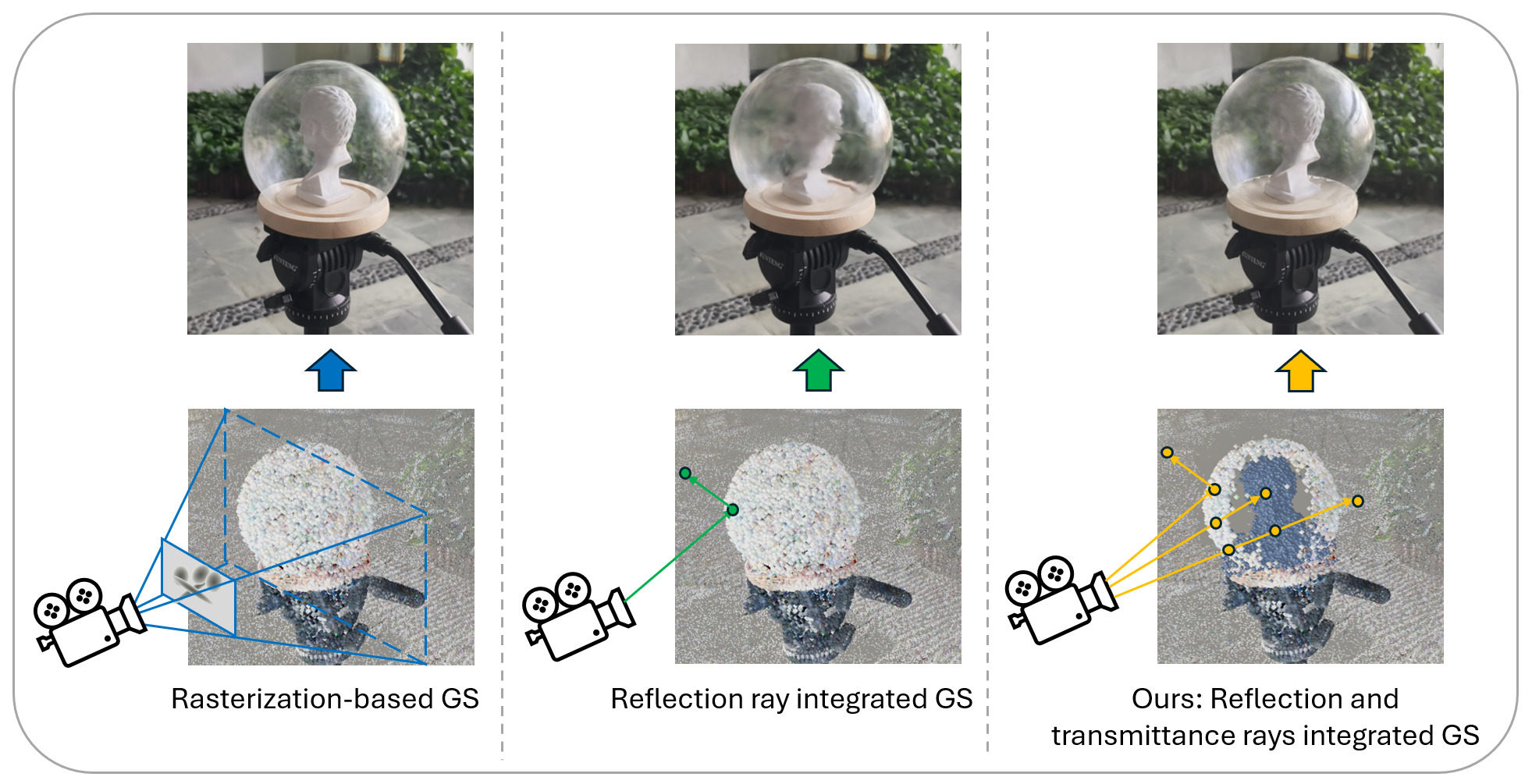}
    \caption{We present a method that optimizes the quality of specular reflection and transmittance concurrently in GS-based method. For the rasterization-based gaussian splatting method \cite{3dgs} in the left image, the specular reflection details are blurred on the surface of objects. For the reflection rays integrated method \cite{envgs} in the middle image, the object behind the semi-transparent surface is blurred. We solve both the complex reflection and transmittance ray effects at the same time by integrating reflection and transmittance rays into gaussian splatting. Then, we can see the transmittance appearance and specular reflection details, as shown in the right image.}
    \label{fig:fig1}
\end{figure}

In this paper, we present a novel framework to further improve the quality of specular reflection and transmittance appearance simultaneously in GS-based methods. We utilize the microfacet model to represent complex material properties on surfaces. We reconstruct the 3D scenes using three separate Gaussian primitives: diffuse Gaussian, reflection Gaussian, and transmittance Gaussian. We start with the rasterization process using diffuse Gaussians to obtain the per-pixel diffuse color, surface position, normal map, roughness map, and reflectance map. The microfacet model is integrated to calculate the bidirectional scattering distribution function (BSDF) on the surfaces of diffuse Gaussians. Then, we generate reflection and transmittance rays, based on the normal map at surface positions, and utilize differentiable ray tracing on reflection and transmittance Gaussians, respectively. We assume that the outer layer of transparent objects is infinitely thin, so transmittance rays maintain the direction of incoming camera rays. The specular reflection color is obtained by ray tracing on reflection Gaussian. The objects 
behind transparent surfaces
are reconstructed by transmittance Gaussians. Transmittance rays exiting from the back of transparent objects continue to intersect with the diffuse Gaussians to reconstruct the light paths of seeing through transparent objects. Finally, we blend diffuse, reflection, and transmittance colors using the BSDF to obtain the final rendering results.

To train diffuse, specular reflection, and transmittance components jointly from input 2D images, we first generate segmentation masks for transparent objects using SAM2 \cite{sam,sam2} guided by GroundingDINO \cite{dino}. Reflection and transmittance rays are traced simultaneously on the surface of transparent objects, while only reflection rays are traced elsewhere. We extract meshes of transparent objects to guide the transmittance rays and limit the transmittance Gaussians inside transparent objects. We add a regularization term on specular strength \cite{materialrefgs} to force the reflection and transmittance colors to have a larger percentage on transparent objects. 

In summary, the technical contributions in this paper are the following:
\begin{itemize}
\item We integrate a microfacet model to reconstruct complex material properties.
\item We introduce separate transmittance Gaussian primitives to model the interiors of transparent objects.
\item We extract the mesh to guide the ray tracing on transmittance Gaussian and use masks of transparent objects to guide the training of material properties.
\end{itemize}

\section{Related Work}

The target of novel view synthesis is to generate images from new viewpoints of a scene based on a limited set of posed images. Neural radiance fields (NeRF)~\cite{NeRF} adopt implicit multilayer perceptrons (MLPs) to reconstruct 3D scenes and leverage differentiable volume rendering to obtain rendering results, achieving a high quality of appearance in novel views. Subsequent methods have further enhanced rendering quality \cite{ref-nerf,nu-nerf} and accelerated both training and rendering speed \cite{instantngp}.

Recently, 3D Gaussian Splatting (3D-GS) \cite{3dgs} achieves real-time rendering and superior rendering quality over NeRF-based methods by rasterizing 3D Gaussian primitives. 2D Gaussian Splatting (2D-GS) \cite{2dgs} flattens 3D Gaussians into 2D Gaussian primitives to achieve better surface alignment. In our work, we adopt 2D Gaussian primitives as the scene representation to reconstruct better surface geometry and normals.

3D-GS and 2D-GS struggle to accurately model specular reflection and transmittance concurrently due to their use of spherical harmonics (SH) to reconstruct appearance, which causes blurry results. Recent approaches \cite{gaussianshader,3dgs-dr} use the environment map to model specular reflection, but they fail to model near-field reflections and transmittance. Other methods \cite{ref-gaussian,materialrefgs} integrate the PBR materials and deferred shading to achieve better physical properties of the surface. However, they only focus on specular reflections and fail to model the transmittance appearance. 
EnvGS \cite{envgs} integrate ray tracing on Gaussians to achieve better specular reflection, but they also fail to model the transmittance effects of transparent objects. TransparentGS \cite{transparentgs} models transparent objects as separate Gaussian primitives and bakes the environment around them for deferred shading. However, it ignores objects behind the transparent surfaces. Our method integrates the microfacet model without the split-sum approximation to reconstruct surface properties. We adopt ray tracing to model specular effects and reconstruct objects behind transparent surfaces.

\begin{figure}[tb]
    \centering
    \includegraphics[width=1\linewidth]{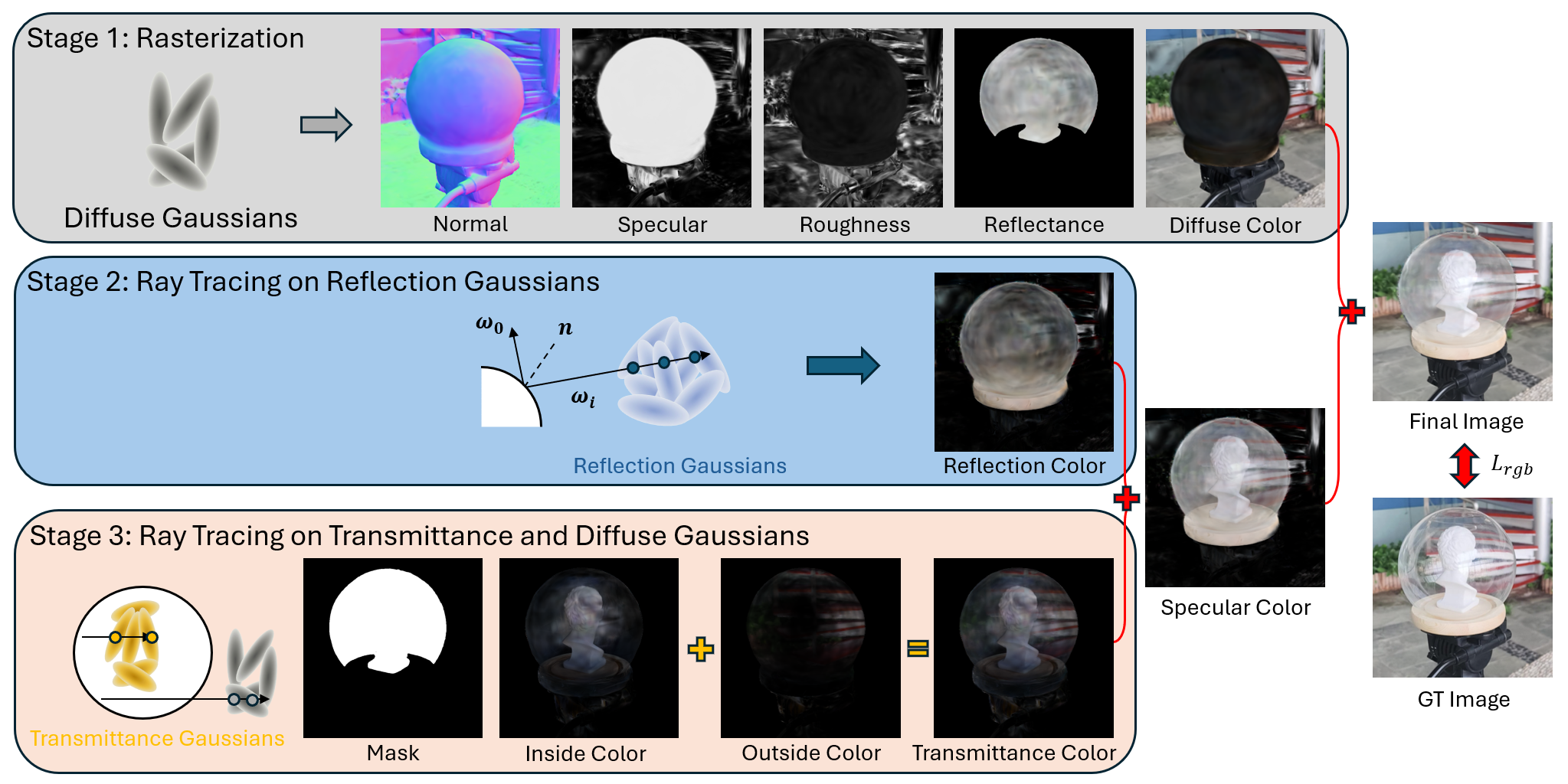}
    \caption{{\bf Overview of our method.} In the first stage, we start with the rasterization process using diffuse Gaussians to obtain per-pixel material maps and diffuse color. In the second stage, we generate specular reflection rays from the surface of diffuse Gaussians and utilize ray tracing on reflection Gaussians. In the third stage, we trace first-bounce transmittance rays from the surface into the transparent objects on transmittance Gaussians to obtain the inside color and second-bounce transmittance rays from the back surface of the mesh on diffuse Gaussian again to obtain the outside color. Finally, we combine the diffuse, reflection and transmittance colors for the final image. We jointly optimize these three types of Gaussians using ground truth images for supervision.}
    \label{fig:fig2}
\end{figure}

\section{Method}


3D-GS \cite{3dgs} represents a scene using differentiable 3D Gaussian primitives defined by a covariance matrix $\sum$ and a position $\mu$ in world space
\begin{equation}
  G(x) = e^{-\frac{1}{2}\,(x)^{\top}\Sigma^{-1}(x)} \;.
\end{equation}
During the rendering process, the 3D Gaussians are projected to 2D splats using the transformation matrix from world space to image space. The rendering color $C$ is computed by $\alpha$-blending the 2D splats in front-to-back order, such that
\begin{equation}
  C = \sum_{i=1}^{N} C_io_i\prod_{j=1}^{i-1}(1-o_j) \;,
\end{equation}
where $o_i$ is the opacity and $C_i$ is the view-dependent color represented by spherical harmonics of each Gaussian.

2D-GS \cite{2dgs} uses 2D Gaussian primitives to represent a scene and provide better surface reconstruction. The 2D Gaussian is defined in a local tangent plane in world space with two principal tangential vectors $\vec{t}_u$ and $\vec{t}_v$ and a scaling vector $(s_u,s_v)$. Because the specular reflection and transmittance require accurate surface representation, we adopt the 2D Gaussian as our base primitive. However, 2D-GS integrates spherical harmonics to model scene appearance, which introduces a blurry appearance for highly view-dependent effects, including specular reflection and transmittance. To address this, we introduce separate Gaussian primitives to model specular effects.

\subsection{Modeling Specular Light Transport}
To reconstruct scenes with both specular reflection and transmittance, we separate the Gaussian primitives into diffuse, reflection, and transmittance Gaussian primitives. The diffuse Gaussian is initialized using the sparse point cloud obtained from Structure-from-Motion (SfM)~\cite{detect,photo}. The reflection and transmittance Gaussians are both randomly initialized within the scene's bounding box \cite{envgs}.

Each Gaussian is integrated into depth $d$, normal $\vec{n}$, roughness $\alpha$, base reflectance $f_0$ and specular blending weight $k_s$ properties. How we use these properties for deferred shading is demonstrated in \autoref{sec:microfacet}. We first use 2D Gaussian splatting to render the diffuse Gaussian. These properties of the diffuse Gaussians are $\alpha$ blended by rasterization to generate per-pixel material maps for deferred shading:
\begin{equation}
  x = \sum_{i=1}^{N} x_io_i\prod_{j=1}^{i-1}(1-o_j), \; x\in\{d, \vec{n}, \alpha, f_0, k_s\} \;,
\end{equation}
where $o_i$ is the opacity of the $i$-th Gaussian primitive.

After we obtain these material maps, we compute surface positions based on the depth map. Then, the reflection direction can be calculated based on the surface position and normal map. The transmittance direction is taken as the direction of the incoming camera rays, assuming infinitely thin semi-transparent surfaces. The reflection and transmittance Gaussians are rendered by a differentiable ray tracer \cite{envgs} with the reflection and transmittance rays to obtain the reflection color $C_r$ and transmittance color $C_t$, respectively. Based on the diffuse color $C_d$ obtained through rasterization on the diffuse Gaussian, we can derive the final color using
\begin{equation}
  C = (1-k_s)C_d + k_s(f_r C_r + f_t C_t) \;.
\end{equation}
where $k_s$ is the blending weight for the specular term, $f_r$ is the bidirectional reflectance distribution function (BRDF) and $f_t$ is the bidirectional transmittance distribution function (BTDF).

\subsection{Microfacet Material Model}
\label{sec:microfacet}

To 
model dielectric surfaces, we incorporate a microfacet material. We divide the rendering color $C$ at the surface position $\vec{x}$ in the viewing direction $\vec{\omega}_o$ into two parts: the diffuse term and the specular term \cite{pbshading}:
\begin{equation}
  C(\vec{x},\vec{\omega}_o) = (1-k_s) C_d(\vec{x}) + k_s L_s(x,\omega_o)
\end{equation}
where $L_s$ is the outgoing specular radiance.

We adopt the rendering equation \cite{renderingequation} to calculate outgoing specular radiance $L_s$ at the surface position $\vec{x}$ in the direction $\vec{\omega}_o$:
\begin{equation}
  L_s(\vec{x},\vec{\omega}_o) = \int_\Omega L_i(\vec{x},\vec{\omega}_i) f_s(\vec{x},\vec{\omega}_i, \vec{\omega}_o)(\omega_i\cdot n) \text{d}\vec{\omega}_i
\end{equation}
where $L_i$ is the incident radiance from the upper hemisphere, $\vec{\omega}_i$ is the incidence direction and $\vec{n}$ is the surface normal. $f_s$ is the bidirectional scattering distribution function (BSDF), which can be composed by the BRDF $f_r$ and the BTDF $f_t$, such that
\begin{equation}
  f_s(\vec{x},\vec{\omega}_i, \vec{\omega}_o) = f_r(\vec{x},\vec{\omega}_i, \vec{\omega}_o) + f_t(\vec{x},\vec{\omega}_i, \vec{\omega}_o)
\end{equation}

For the BRDF $f_r$, we adopt the Torrance-Sparrow BRDF \cite{cook-torrance}
\begin{equation}
  f_r(x,\omega_i, \omega_o) = \frac{DGF}{4(\vec{\omega}_o\cdot \vec{n})(\vec{\omega}_i\cdot \vec{n})}
  \label{eq:brdf}
\end{equation}
where $D$, $G$, $F$ denote the normal distribution function, the shadowing-masking term and the Fresnel term, respectively.

For the normal distribution function, we use the isotropic Trowbridge–Reitz model \cite{ggx}
\begin{equation}
  D(\vec{n},\vec{\omega}_h,\alpha) = \frac{\alpha^2}{\pi\Big((\vec{n}\cdot \vec{\omega}_h)^2(\alpha^2-1)+1\Big)^2}
\end{equation}
where $\vec{\omega}_h$ is the half vector between the incidence direction $\vec{\omega}_i$ and the outgoing direction $\vec{\omega}_o$ such that ${\vec{\omega}_h=(\vec{\omega}_i+\vec{\omega}_o)/\|\vec{\omega}_i+\vec{\omega}_o\|}$, and $\alpha$ is the roughness.

When $\alpha$ approaches zero, it indicates near-perfect specular reflection. During training, the roughness approaching zero causes the normal distribution term to become excessively large, thereby adversely affecting the quality of training. Therefore, we apply a mapping from such roughness values to input $\alpha$ values \cite{pbrt}:
\begin{multline}
  g(\alpha) = 1.62142 + 0.819955\ln{\alpha} + 0.1734(\ln{\alpha})^2 + 0.0171201(\ln{\alpha})^3 \\
  + 0.000640711(\ln{\alpha})^4
\end{multline}

For the BTDF $f_t$, we consider specular transmission:
\begin{equation}
  f_t(\vec{x},\vec{\omega}_i, \vec{\omega}_o) = (1-F)\frac{\delta(\vec{\omega}_i-\vec{\omega}_t)}{(\vec{\omega}_i\cdot \vec{n})}
\end{equation}\
where $\vec{\omega}_t$ is the specular transmission direction and $\delta$ is the Dirac delta function.

\begin{figure}[tb]
    \centering
    \includegraphics[width=1.0\linewidth]{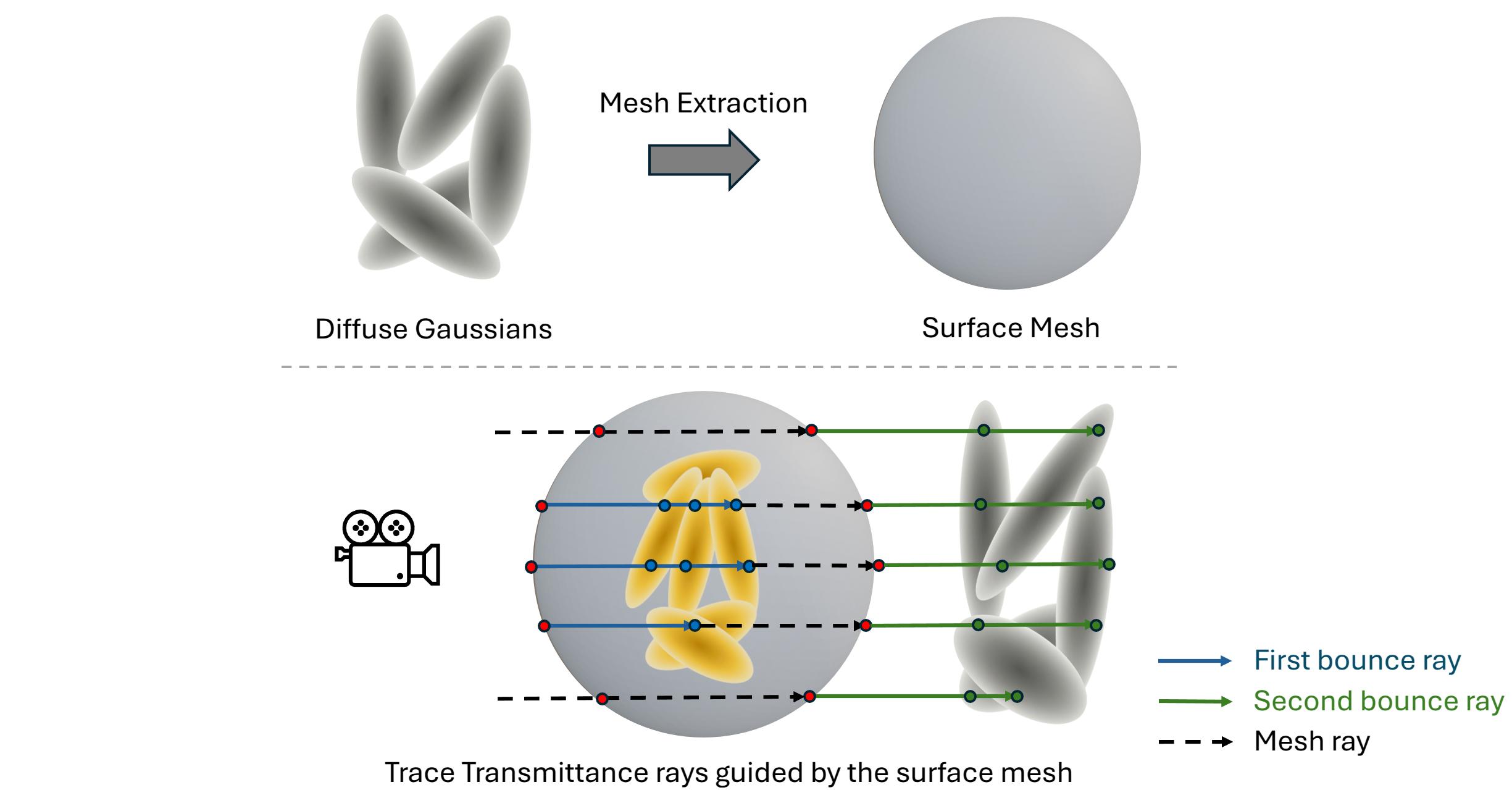}
    \caption{{\bf Mesh-guided ray tracing for transparent objects.} The surface mesh is extracted from the diffuse Gaussian. The first bounces of transmittance rays are traced from the depth positions of the diffuse Gaussian on the transmittance Gaussian. The second bounces of transmittance rays are traced from the back surface positions of the extracted mesh on the diffuse Gaussian. To obtain the back surface positions, we trace rays on the mesh to get the second nearest depth.}
    \label{fig:fig3}
\end{figure}

\subsection{Optimizations}
We employ a multi-stage reconstruction process for these three different Gaussian primitives. \autoref{fig:fig2} illustrates the overview of our rendering pipeline.

In the first stage, we obtain the material maps $\{d, \vec{n}, \alpha, f_0, k_s\}$ and the diffuse color $C_d$ based on the diffuse Gaussian. 

In the second stage, we add specular reflection rays starting from the surface positions of the diffuse Gaussian to obtain the reflection color $C_r$ from the reflection Gaussian. We adopt the joint optimization of the diffuse Gaussian and the reflection Gaussian to obtain better material quality and extract high-quality surface meshes. 

In the final stage, after obtaining high-quality surface positions and normals, we first extract the surface mesh from the diffuse Gaussian. Then, we trace the transmittance rays guided by the extracted mesh to obtain the transmittance color $C_t$ and jointly optimize these three Gaussian primitives. \autoref{fig:fig3} illustrates the process of mesh-guided ray tracing for transparent objects. Specifically, the first bounce of transmittance rays starts from the surface positions of the diffuse Gaussian and is traced on the transmittance Gaussian. To obtain the color of the external environment as seen through transparent objects, we need to trace rays from the back surface of transparent objects on the diffuse Gaussian again, because the diffuse Gaussians have already been well trained. However, the back surface positions cannot be obtained by the rasterization of the diffuse Gaussian, as we can only get the positions of the nearest depth. Therefore, we trace rays from cameras on the extracted surface mesh to obtain the nearest depth and second nearest depth. The second bounce of transmittance rays starts from the second-nearest surface positions and is traced on the diffuse Gaussians again.

To add transmittance rays only to the parts of transparent objects, we generate segmentation masks for transparent objects using SAM2 \cite{sam,sam2} guided by GroundingDINO \cite{dino}. Based on the observation that surfaces on transparent objects have a high specular strength, we add regularization to the specular map to encourage the specular colors $C_r$ and $C_t$ to dominate the final color for transparent objects. Inspired by MaterialRefGS \cite{materialrefgs}, the masks are used to add the constraint to the specular map $k_s$ at a high target value, such that
\begin{equation}
  \mathcal{L}_\text{spec}=\frac{1}{N_p}\sum_{i=1}^{N_p}\Gamma_i\cdot \text{ReLU}(K_0-k_{s,i}) \;,
\end{equation}
where $N_p$ is the number of pixels in the image, $\Gamma_i$ is the binary mask for the transparent object at pixel $i$, $K_0$ is the target value, and $\operatorname{ReLU}$ is the rectified linear unit \cite{relu}.

\begin{figure}[tb]
    \centering
    \includegraphics[width=0.7\linewidth]{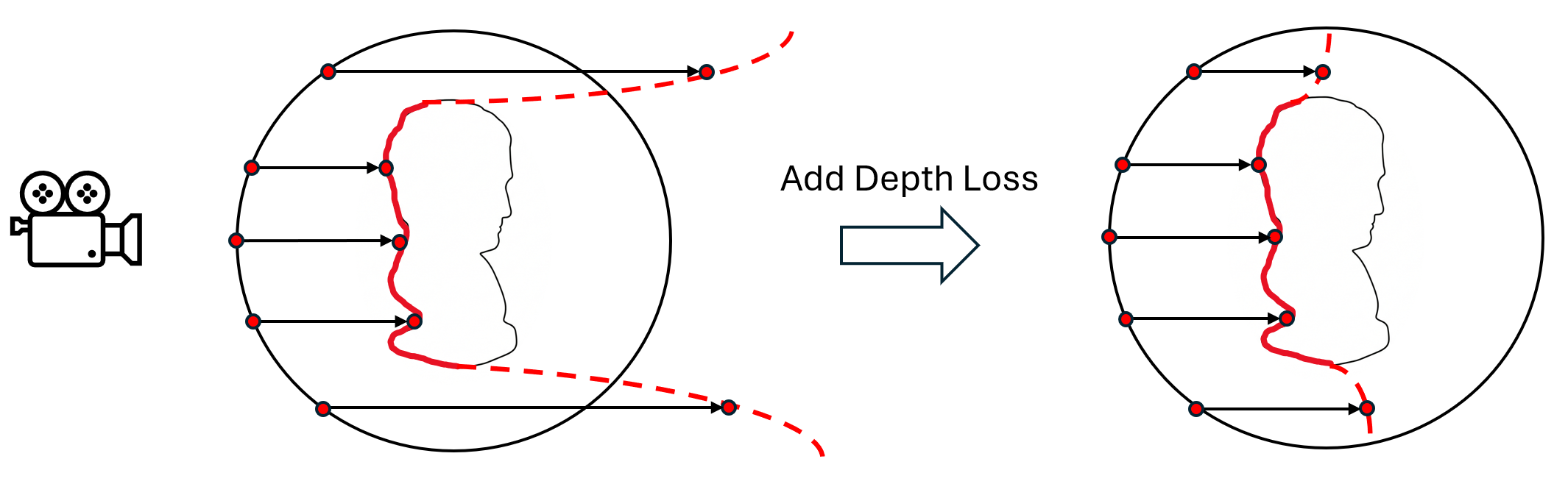}
    \caption{{\bf Illustration of the regularization to the depth.} When reconstructing the internal object, the depth of first-bounce transmittance rays of the part of the object that exists is reconstructed well, as demonstrated by the solid red curve. But in the parts where no object is existing, the depth is trained to the outside the transparent object, as shown by the red dashes curve, because it tries to reconstruct the outside color, which should be rendered by second-bounce transmittance rays. Therefore, we add regularization to the depth to push it back into the transparent object.}
    \label{fig:fig4}
\end{figure}

The transmittance Gaussian is used to reconstruct the objects inside transparent objects, so it needs to be constrained within transparent objects. As illustrated in \autoref{fig:fig4}, we add regularization to the depth map $d$ obtained by the first bounce of transmittance rays, such that
\begin{equation}
  \mathcal{L}_{d}=\frac{1}{N_p}\sum_{i=1}^{N_p}\Gamma_i\cdot \text{ReLU}(d_i-D_i) \;,
\end{equation}
where $D_i$ is the second-nearest depth of the diffuse Gaussian at the $i$-th pixel, specifically on the back surface of transparent objects.

We adopt the constraint for aligning the normal map values $\vec{n}_i$ with the gradients of the depth maps $\vec{N}_d$ \cite{2dgs}
\begin{equation}
  \mathcal{L}_\text{norm}=\frac{1}{N_p}\sum_{i=1}^{N_p}(1-\vec{n}_i 
  \cdot \vec{N}_d) \;,
\end{equation}
where $\vec{N}_d$ is computed with finite differences from nearby pixel positions $\vec{p}_d$ in the depth map, using
\begin{equation}
  \vec{N}_d = \frac{\nabla_x \vec{p}_d \times \nabla_y \vec{p}_d}
{\left\lVert \nabla_x \vec{p}_d \times \nabla_y \vec{p}_d \right\rVert} \;.
\end{equation}

We follow EnvGS \cite{envgs} to use monocular normal estimates $\vec{N}_m$ to supervise the normal map values $\vec{n}_i$ and perceptual loss \cite{envgs} to enhance the perceived quality of the rendered image:
\begin{align}
  \mathcal{L}_\text{mono} &= \frac{1}{N_p}\sum_{i=1}^{N_p}(1-\vec{n}_i 
  \cdot \vec{N}_m)\\
  \mathcal{L}_\text{perc} &= \left\lVert \Phi(\mathbf{I}) - \Phi(\mathbf{I}_{gt}) \right\rVert_{1} \;,
\end{align}
where $\Phi$ is the pre-trained VGG-16 network \cite{vgg}, $\mathbf{I}$ is the rendered image and $\mathbf{I}_{gt}$ is the ground truth image.

The final loss function is a weighted combination of these loss terms, such that
\begin{equation}
  \mathcal{L}=\mathcal{L}_\text{rgb} + \lambda_\text{spec}\mathcal{L}_\text{spec} + \lambda_d\mathcal{L}_{d} + \lambda_\text{norm}\mathcal{L}_\text{norm} + \lambda_\text{mono}\mathcal{L}_\text{mono} + \lambda_\text{perc}\mathcal{L}_\text{perc} \;,
\end{equation}
where $\mathcal{L}_\text{rgb}$ is the photometric reconstruction loss combining the $\mathcal{L}_{1}$ with the D-SSIM term\cite{3dgs}, and $\lambda$ values are the weights. We use ${\lambda_\text{spec} = 0.2}$, ${\lambda_d = 0.2}$, ${\lambda_\text{norm} = 0.04}$, ${\lambda_\text{mono} = 0.01}$, ${\lambda_\text{perc} = 0.01}$ for all experiments in this paper.

\section{Implementation and Results}
Our training process is divided into 3 stages. We adopt rasterization of diffuse Gaussians for 3,000 steps to obtain all material maps. Then, we generate specular reflection rays and utilize the differentiable ray tracing framework from EnvGS \cite{envgs} to reconstruct reflection colors for 17,000 steps. We extract surface mesh of diffuse Gaussians at step 20,000 using the truncated signed distance fusion (TSDF) from 2DGS \cite{2dgs}. After that, the mesh is fixed and we generate transmittance rays guided by the extracted mesh to obtain transmittance colors and jointly optimize these three types of Gaussians for 40,000 steps. The regularization $\mathcal{L}_{d}$ is added at step 40,000 to push the depth back into the transparent surfaces. The target value $K_0$ of the constraint to the specular map is set to 0.9.

\subsection{Datasets and Comparisons}
We evaluate our method on datasets that contain complex specular reflections and transparent objects in real-world scenes to demonstrate the ability to reconstruct complex specular reflections and transmittance appearance concurrently. Specifically, for reflection scenes, we evaluate three scenes from Ref-Real \cite{ref-nerf}: Sedan, Spheres and Toycar. For transparent scenes, we evaluate four scenes from NU-NeRF \cite{nu-nerf}: BallStatue, PopcornCup, RealBottle and TallBottle. We use the standard PSNR, SSIM \cite{ssim} and LPIPS \cite{lpips} metrics to evaluate the quality.

We compare our method with the following baselines: 3D-GS \cite{3dgs}, GaussianShader \cite{gaussianshader}, Ref-GS \cite{refGS}, Ref-Gaussian \cite{ref-gaussian} and EnvGS \cite{envgs}. 3D-GS are designed for general diffuse appearance and others are the state-of-the-art (SOTA) methods for specular ray effects. For transparent scenes, we add NU-NeRF \cite{nu-nerf} for comparison, because it is the SOTA method for transparent objects based on neural signed distance field.

\begin{figure}[tb]
    \centering
    \includegraphics[width=1.0\linewidth]{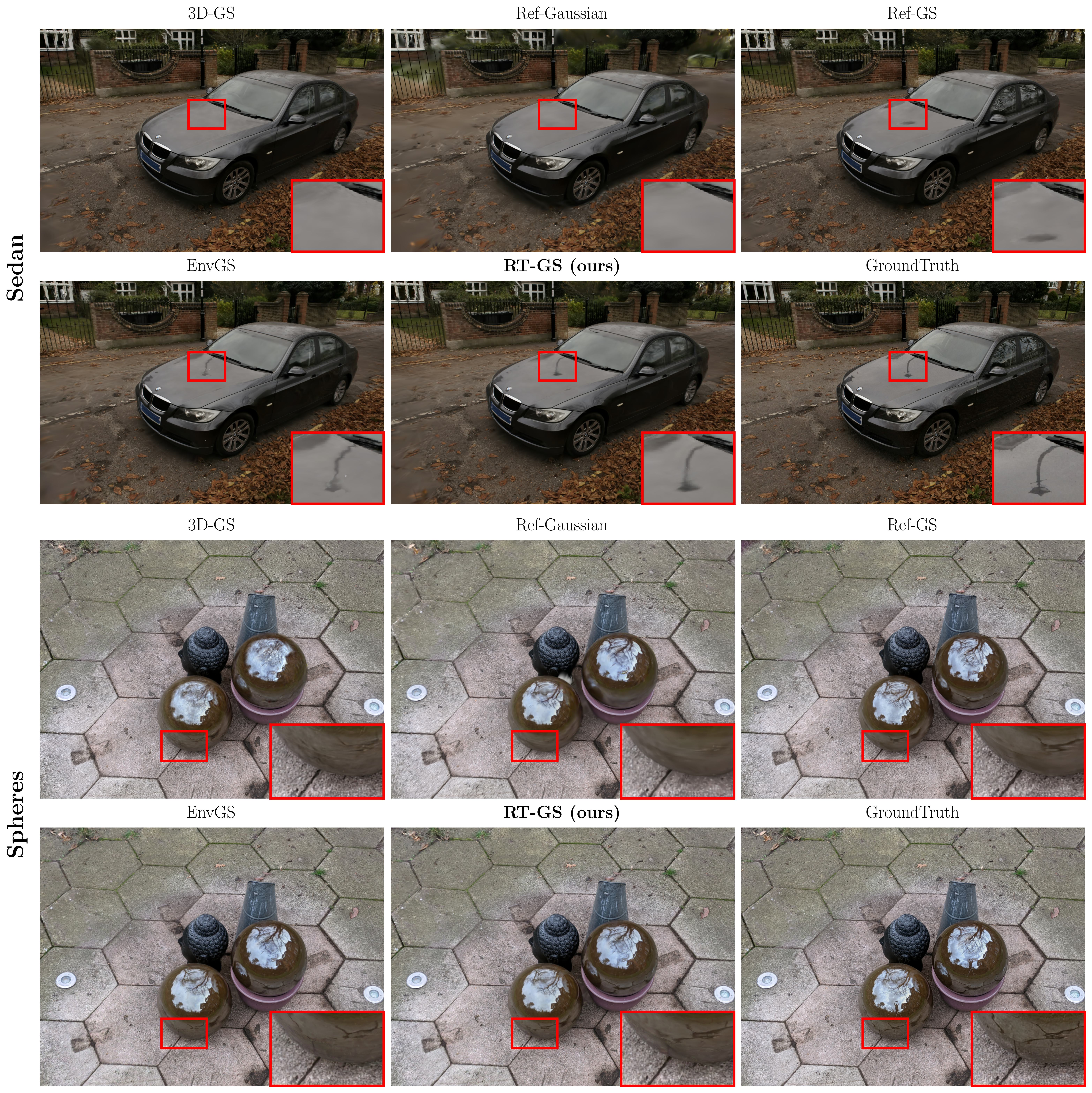}
    \caption{{\bf Qualitative comparison on Ref-Real \cite{ref-nerf} Scenes.}}
    \label{fig:comp1}
\end{figure}

\definecolor{best}{HTML}{F6C6C6}    
\definecolor{second}{HTML}{F8D7B9}  
\definecolor{third}{HTML}{FFF2B3}   

\newcommand{\Best}[1]{\cellcolor{best}{#1}}
\newcommand{\Second}[1]{\cellcolor{second}{#1}}
\newcommand{\Third}[1]{\cellcolor{third}{#1}}

\begin{table}[tb]
  \caption{{\bf Quantitative comparison on the Ref-Real \cite{ref-nerf} dataset.} Best results are highlighted as
  \colorbox{best}{\strut 1st}, \colorbox{second}{\strut 2nd}, \colorbox{third}{\strut 3rd}.}
  \label{tab:table1}
  \centering
  \footnotesize
  \setlength{\tabcolsep}{4pt}
  \renewcommand{\arraystretch}{1.12}

  \begin{adjustbox}{max width=\linewidth, center}
  \begin{tabular}{@{}l|ccc|ccc|ccc@{}}
    \toprule
    \multirow{2}{*}{Method}
      & \multicolumn{3}{c|}{Sedan}
      & \multicolumn{3}{c|}{Spheres}
      & \multicolumn{3}{c}{Toycar} \\
    \cmidrule(lr){2-4}\cmidrule(lr){5-7}\cmidrule(lr){8-10}
      & PSNR$\uparrow$ & SSIM$\uparrow$ & LPIPS$\downarrow$
      & PSNR$\uparrow$ & SSIM$\uparrow$ & LPIPS$\downarrow$
      & PSNR$\uparrow$ & SSIM$\uparrow$ & LPIPS$\downarrow$ \\
    \midrule

    3DGS \cite{3dgs}
      & \Third{25.556} & \Best{0.729} & \Best{0.294}
      & 22.386 & 0.590 & \Third{0.239}
      & \Third{24.394} & \Third{0.661} & \Second{0.229} \\

    GaussianShader \cite{gaussianshader}
      & 22.759 & 0.648 & 0.395
      & 22.080 & 0.599 & 0.268
      & 23.641 & \Third{0.661} & 0.283 \\

    Ref-GS \cite{refGS}
      & 25.522 & 0.724 & 0.338
      & 22.236 & 0.593 & 0.274
      & 24.237 & 0.646 & 0.270 \\

    Ref-Gaussian \cite{ref-gaussian}
      & 24.324 & 0.675 & 0.380
      & \Third{22.788} & \Third{0.614} & 0.278
      & 23.991 & 0.644 & 0.279 \\

    EnvGS \cite{envgs}
      & \Second{26.142} & \Third{0.727} & \Third{0.300}
      & \Second{23.031} & \Best{0.620} & \Best{0.196}
      & \Best{24.759} & \Best{0.669} & \Best{0.206} \\

    \textbf{RT-GS (ours)}
      & \Best{26.203} & \Second{0.728} & \Second{0.298}
      & \Best{23.050} & \Best{0.620} & \Second{0.198}
      & \Second{24.693} & \Second{0.668} & \Third{0.231} \\

    \bottomrule
  \end{tabular}
  \end{adjustbox}
\end{table}

\subsubsection{Comparisons on Reflection Datasets.}
\autoref{fig:comp1} shows the qualitative comparison on the Ref-Real \cite{ref-nerf} dataset. Our method provides better reflection details on the specular surfaces, such as the streetlight reflected on the hood and near-field reflections on the sphere. \autoref{tab:table1} shows the quantitative results, demonstrating that our method achieves the best performance on all other baselines and competitive results compared to EnvGS.

\begin{figure}[tb]
    \centering
    \includegraphics[width=0.91\linewidth]{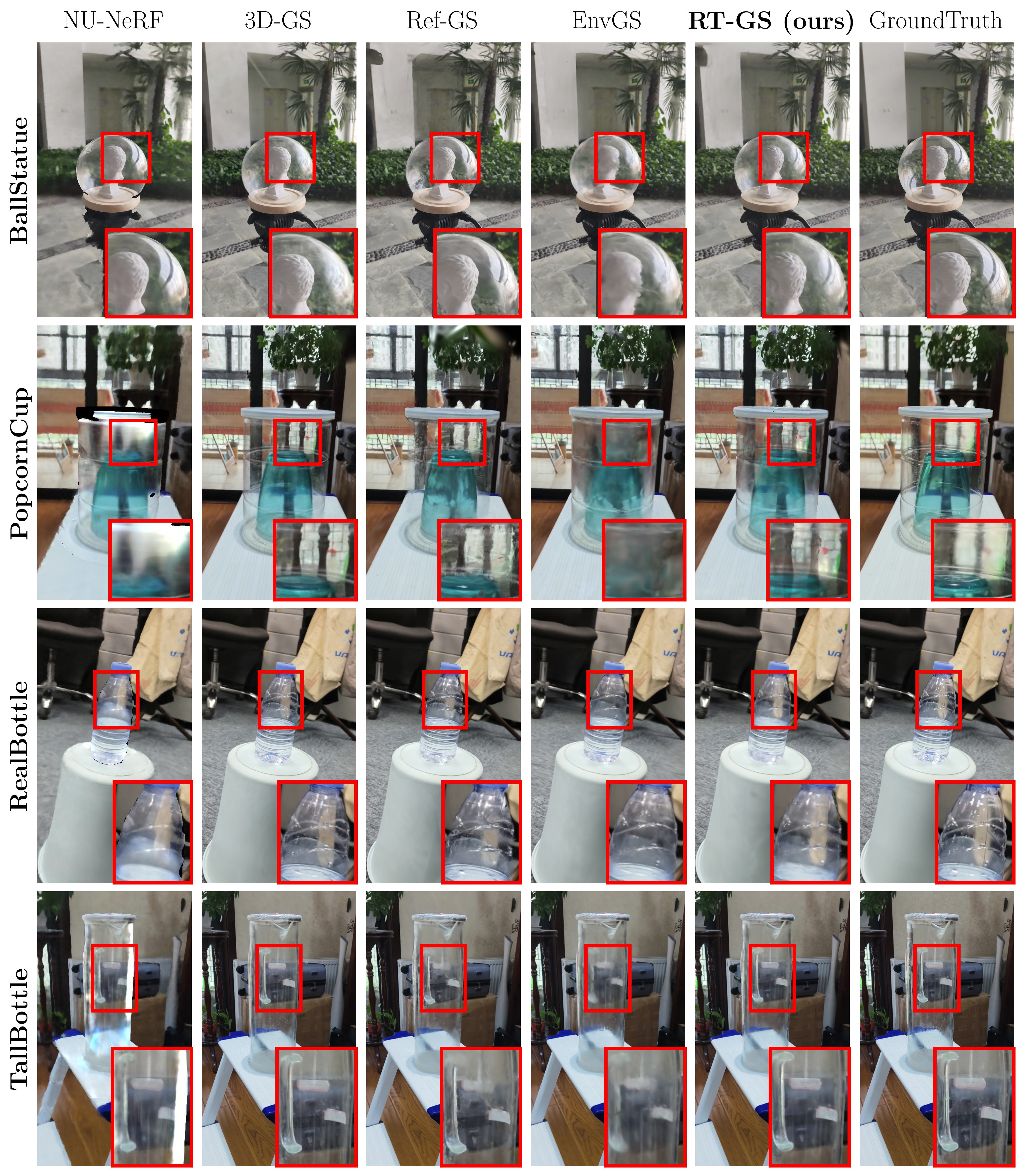}
    \caption{{\bf Qualitative comparison on NU-NeRF \cite{nu-nerf} Scenes.}}
    \label{fig:comp2}
\end{figure}

\begin{table}[tb]
  \caption{{\bf Quantitative comparison on NU-NeRF \cite{nu-nerf} scenes.} Best results are highlighted as
  \colorbox{best}{\strut 1st}, \colorbox{second}{\strut 2nd}, \colorbox{third}{\strut 3rd}.}
  \label{tab:table2}
  \centering
  \footnotesize
  \setlength{\tabcolsep}{3.5pt}
  \renewcommand{\arraystretch}{1.12}

  \begin{adjustbox}{max width=\linewidth, center}
  \begin{tabular}{@{}l|ccc|ccc|ccc|ccc@{}}
    \toprule
    \multirow{2}{*}{Method}
      & \multicolumn{3}{c|}{BallStatue}
      & \multicolumn{3}{c|}{PopcornCup}
      & \multicolumn{3}{c|}{RealBottle}
      & \multicolumn{3}{c}{TallBottle} \\
    \cmidrule(lr){2-4}\cmidrule(lr){5-7}\cmidrule(lr){8-10}\cmidrule(lr){11-13}
      & PSNR$\uparrow$ & SSIM$\uparrow$ & LPIPS$\downarrow$
      & PSNR$\uparrow$ & SSIM$\uparrow$ & LPIPS$\downarrow$
      & PSNR$\uparrow$ & SSIM$\uparrow$ & LPIPS$\downarrow$
      & PSNR$\uparrow$ & SSIM$\uparrow$ & LPIPS$\downarrow$ \\
    \midrule

    NU-NeRF \cite{nu-nerf}
      & 23.453 & 0.706 & 0.345
      & 17.031 & 0.727 & 0.333
      & 25.818 & 0.722 & 0.336
      & 18.923 & 0.707 & 0.344 \\

    3DGS \cite{3dgs}
      & \Best{27.303} & \Best{0.899} & \Best{0.142}
      & \Second{25.679} & \Second{0.869} & \Second{0.202}
      & \Best{30.740} & \Best{0.936} & \Best{0.162}
      & \Second{25.455} & \Third{0.883} & \Second{0.173} \\

    GaussianShader \cite{gaussianshader}
      & \Third{25.589} & 0.872 & 0.206
      & 16.618 & 0.701 & 0.417
      & 22.610 & 0.824 & 0.326
      & 18.137 & 0.734 & 0.389 \\

    Ref-GS \cite{refGS}
      & 23.891 & 0.856 & 0.263
      & 22.919 & 0.817 & 0.342
      & 29.527 & \Second{0.926} & 0.249
      & 24.818 & 0.860 & 0.267 \\

    Ref-Gaussian \cite{ref-gaussian}
      & 15.138 & 0.660 & 0.479
      & 22.837 & 0.829 & 0.279
      & 28.019 & 0.896 & 0.236
      & 23.766 & 0.845 & 0.261 \\

    EnvGS \cite{envgs}
      & 25.508 & \Third{0.878} & \Third{0.182}
      & \Third{23.898} & \Third{0.847} & \Third{0.232}
      & \Third{30.244} & 0.915 & \Third{0.178}
      & \Third{25.393} & \Second{0.886} & \Third{0.183} \\

    \textbf{RT-GS (ours)}
      & \Second{26.053} & \Second{0.885} & \Second{0.174}
      & \Best{25.720} & \Best{0.874} & \Best{0.199}
      & \Second{30.298} & \Third{0.918} & \Second{0.174}
      & \Best{26.225} & \Best{0.888} & \Best{0.172} \\

    \bottomrule
  \end{tabular}
  \end{adjustbox}
\end{table}

\subsubsection{Comparisons on Transparent Datasets.}
\autoref{fig:comp2} shows the qualitative comparison on the transparent dataset. Our method accurately reconstructs transmittance appearance and models specular reflections of the transparent object, while other methods cannot achieve both at the same time. Like 3D-GS, it can handle the transmittance appearance but is not good at specular reflections. \autoref{tab:table2} shows the quantitative results, where our method achieves the best results on all other baselines working on reflections and competitive results compared to 3D-GS across all transparent scenes.

\begin{figure}[tb]
    \centering
    \includegraphics[width=1\linewidth]{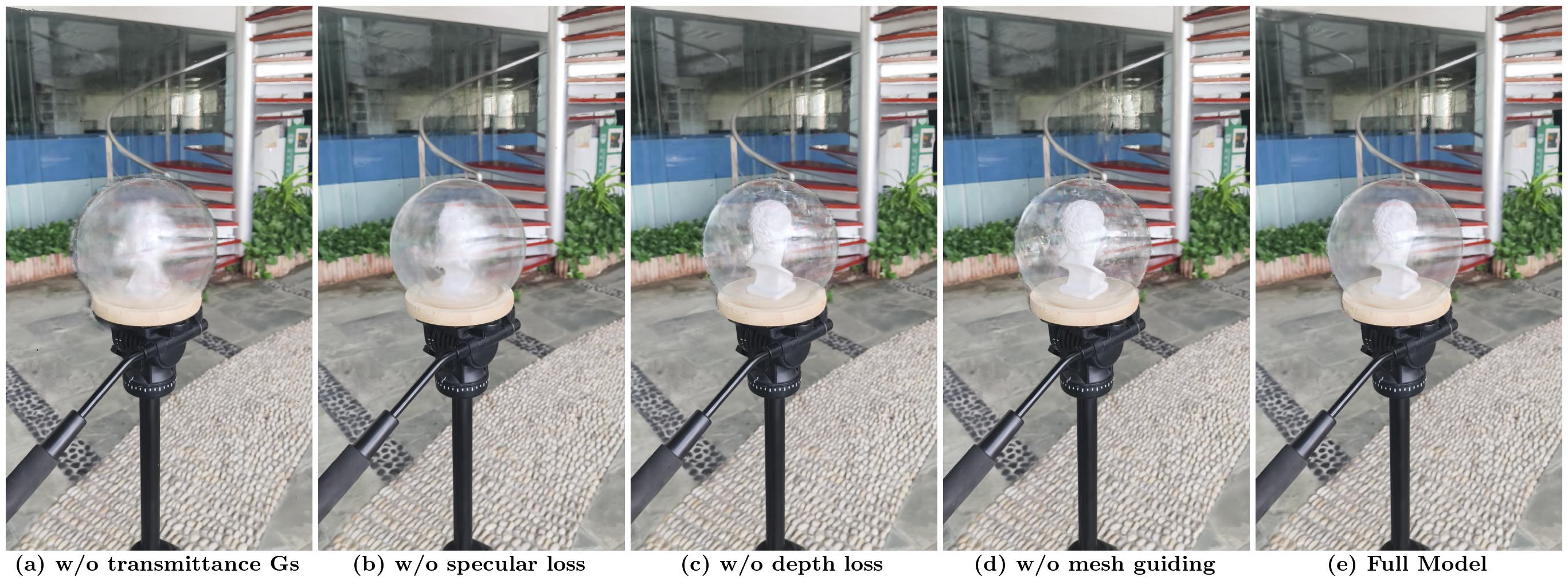}
    \caption{{\bf Ablation study on the BallStatue scene.}}
    \label{fig:ablation1}
\end{figure}

\begin{figure}[tb]
    \centering
    \includegraphics[width=1\linewidth]{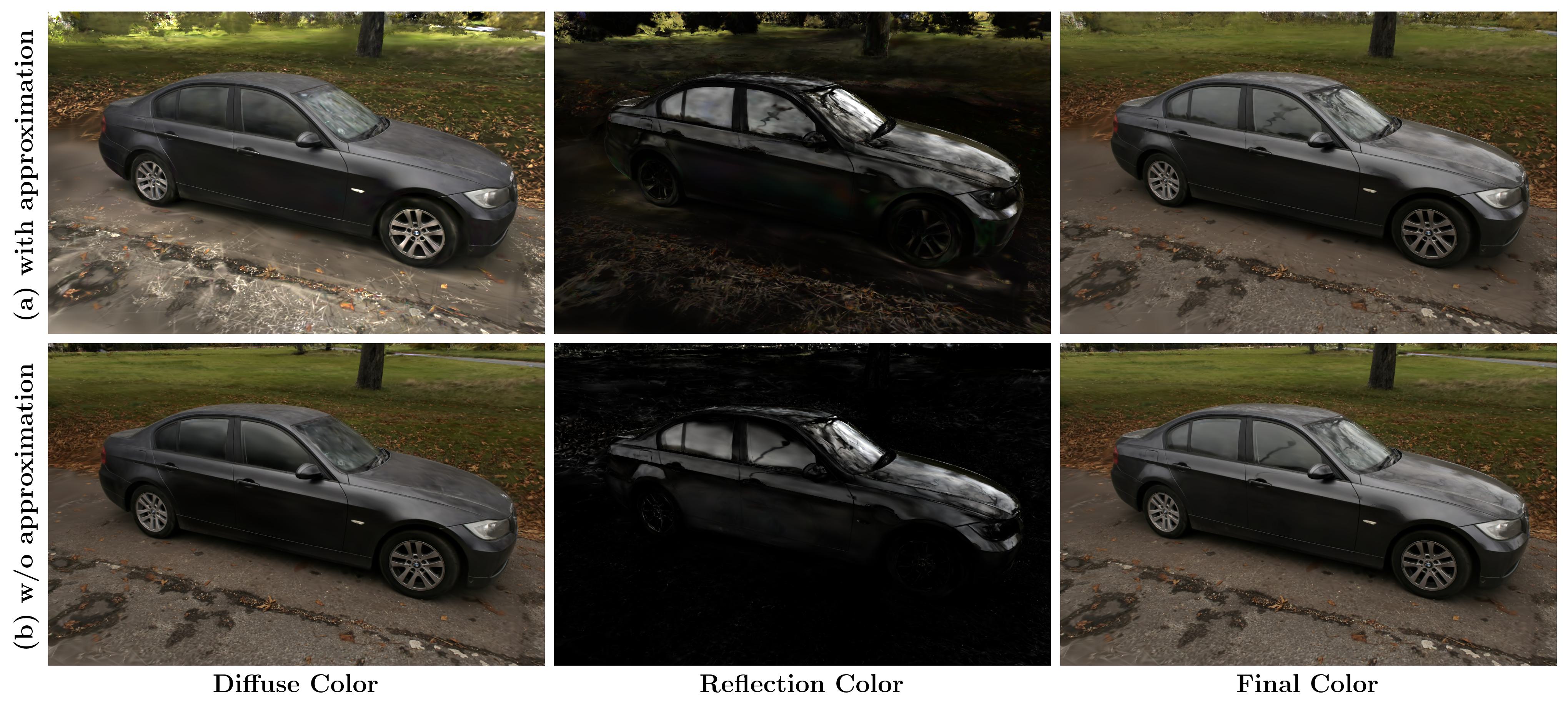}
    \caption{{\bf Ablation study on the Sedan scene.} (a)~Approximating the BSDF with separate components \cite{ref-gaussian, extract} leads to substantial errors in the reconstructions of reflections. (b)~Our solution of using the full microfacet model produces better reflections.}
    \label{fig:ablation2}
\end{figure}

\begin{table}[tb]
  \caption{{\bf Ablation study on the BallStatue scene.}}
  \label{tab:ablation}
  \centering
  \small
  \setlength{\tabcolsep}{10pt}
  \renewcommand{\arraystretch}{1.05}

  \begin{tabular}{@{}lccc@{}}
    \toprule
     & PSNR$\uparrow$ & SSIM$\uparrow$ & LPIPS$\downarrow$ \\
    \midrule
    w/o transmittance Gaussians & 24.844 & 0.866 & 0.196 \\
    w/o specular loss        & 25.673 & 0.882 & 0.174 \\
    w/o depth loss           & 25.941 & 0.881 & 0.172 \\
    w/o mesh guided ray tracing         & 25.889 & 0.880 & 0.174 \\
    \midrule
    Full Model               & \textbf{26.053} & \textbf{0.885} & \textbf{0.174} \\
    \bottomrule
  \end{tabular}
\end{table}

\subsection{Ablation Studies}
We present ablation studies on our method to validate the effect of key algorithmic choices. The quantitative and qualitative comparison results are shown in \autoref{tab:ablation}, \autoref{fig:ablation1}, and \autoref{fig:ablation2} respectively.

\subsubsection{Separate transmittance Gaussians.}
The ``w/o transmittance Gaussians'' option removes the separate transmittance Gaussian and directly utilizes the diffuse Gaussian to reconstruct the object inside the transparent object. As shown in \autoref{fig:ablation1}(a), it fails to reconstruct the internal object.

\subsubsection{Specular map constraint.}
The ``w/o specular loss'' option removes the constraint $\mathcal{L}_\text{spec}$ to the specular map. Without this constraint, the diffuse Gaussian attempts to reconstruct the internal object, which degrades the training process of the transmittance Gaussian. As shown in \autoref{fig:ablation1}(b), the internal object is also failed to be reconstructed.

\subsubsection{Depth map constraint.}
The ``w/o depth loss'' option removes the constraint $\mathcal{L}_{d}$ to the depth map obtained by the first bounce of transmittance rays. Without this constraint, the transmittance Gaussian attempts to model the outside color in areas where there are no internal objects, which is failed to be reconstructed in multiple views. A low quality first bounce of transmittance rays affects the quality of transmittance colors. As shown in \autoref{fig:ablation1}(c), there are artifacts within the transparent object.

\subsubsection{Mesh-guided ray tracing.}
The ``w/o mesh guiding'' option removes the guide of the extracted mesh for tracing transmittance rays. The second bounce of transmittance rays starts from the depth of the first bounce of transmittance rays. Then the depth of the first bounce is free of constraints, like the situation in the "w/o depth loss". Therefore, artifacts show up in \autoref{fig:ablation1}(d).

\subsubsection{Split-sum approximation.}
We also present ablation studies on the split-sum approximation in the microfacet model. The \autoref{fig:ablation2} shows that the split-sum approximation degrades the joint optimization of diffuse and reflection Gaussians. The "with approximation" and "w/o approximation" indicate whether to use or not to use the split-sum approximation in calculating the BRDF term \autoref{eq:brdf} as shown in \autoref{fig:ablation2}(a) and (b), respectively. The "Final Color" is the blended rendering result combining the "Diffuse Color" with "Reflection Color". In \autoref{fig:ablation2}(a), while the car window can produce specular reflections, the diffuse surface is not well trained, like the grass in the background. This is because the blending weight $k_s$ is not trained well during the joint optimization, and the specular term of the ground takes a more dominant role, resulting in a poor effect on the diffuse ground. In \autoref{fig:ablation2}(b), the diffuse and specular reflection terms can be separated well to get both good reflections on the glass and diffuse colors on the ground.

\section{Conclusion and Discussion}
We have introduced RT-GS, a unified framework for reconstructing high-quality specular reflections and the appearance of objects observed through transparent surfaces. Our method integrates a microfacet model for complex material properties and represents the scene using separate Gaussian primitives for diffuse, reflection, and transmittance components. We have also introduced mesh-guided ray tracing and regularization terms for better reconstruction of complex transmittance effects. Extensive experiments on multiple datasets and ablation studies demonstrate the improvements of our method over existing methods in quantitative metrics and visual quality.

\subsubsection{Limitations and future work.}
Our method assumes that the thickness of the outer layer of transparent objects is nearly zero, which makes it difficult to handle the case where the light is strongly curved. Future work could be extend our method to reconstruct thicker transparent objects.

\section*{Acknowledgements}
We thank Feng Liang for comments and discussions. We thank the support of the Berkeley AI Research (BAIR) Lab in providing computational resources.

%
%
\bibliographystyle{splncs04}
\bibliography{main}

\clearpage
\appendix

\begin{figure}[h]
    \centering
    \includegraphics[width=0.72\linewidth]{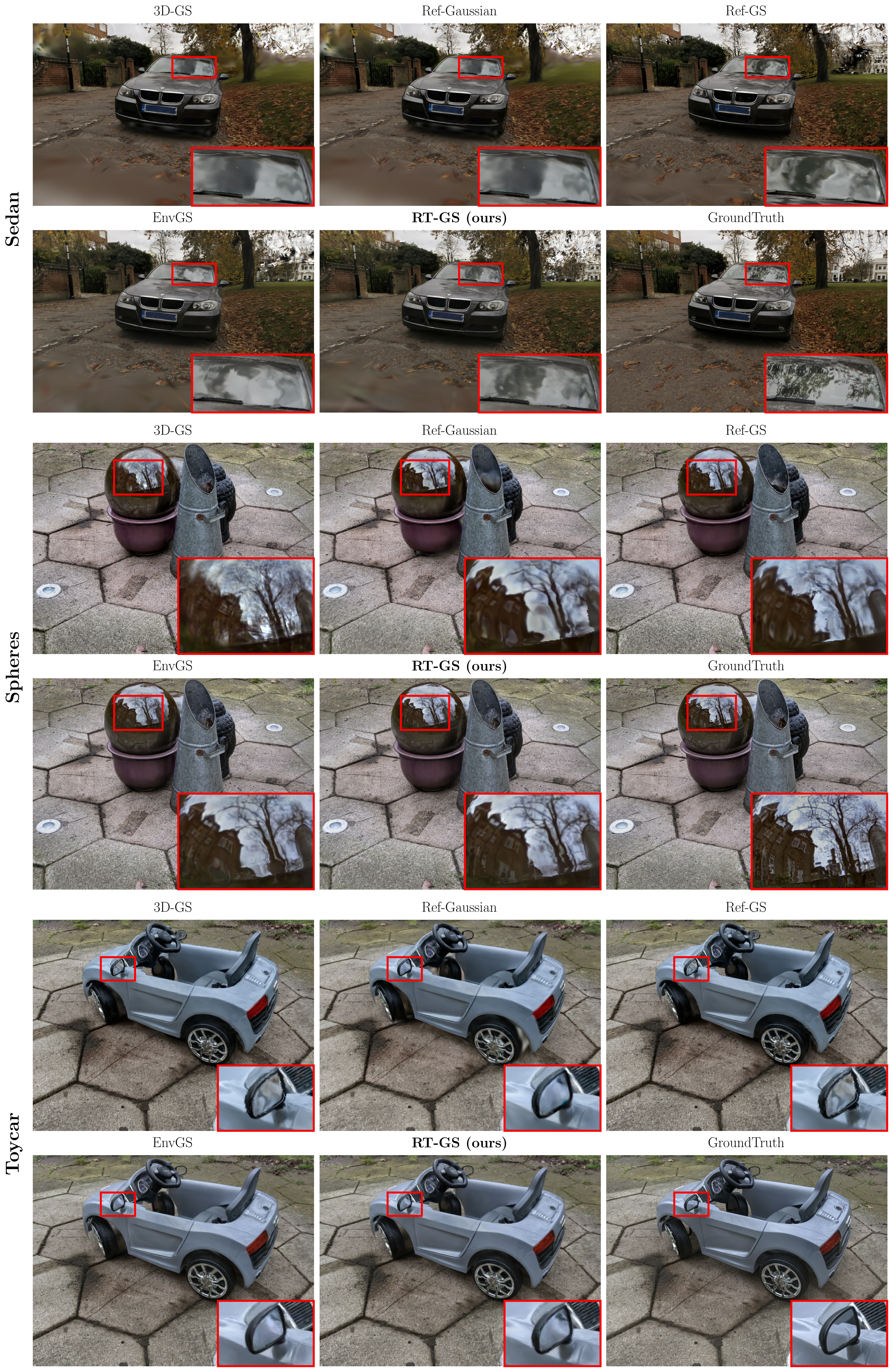}
    \caption{{\bf Qualitative comparison on Ref-Real \cite{ref-nerf} Scenes.}}
    \label{fig:supp1}
\end{figure}

\begin{figure}[tb]
    \centering
    \includegraphics[width=1.0\linewidth]{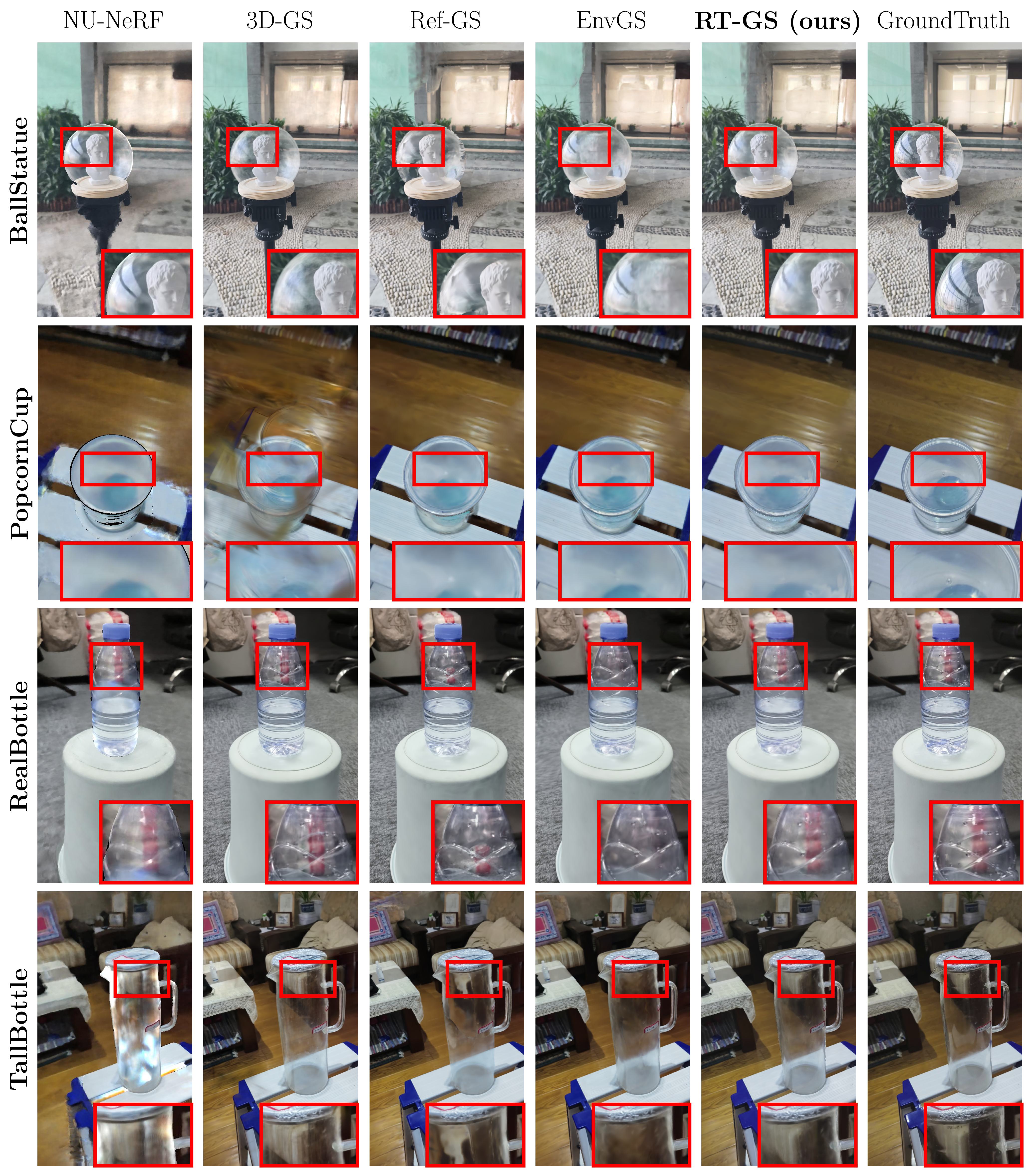}
    \caption{{\bf Qualitative comparison on NU-NeRF \cite{nu-nerf} Scenes.}}
    \label{fig:supp2}
\end{figure}

This appendix presents additional qualitative comparisons on the Ref-Real \cite{ref-nerf} and NU-NeRF \cite{nu-nerf} datasets from different viewpoints. As shown in \autoref{fig:supp1}, our method significantly provides more specular reflection details on specular surfaces. As shown in \autoref{fig:supp2}, our method provides better transmittance appearance results compared to methods that focus on solving reflection problems, while also providing better reflection results compared to other methods.

\end{document}